\documentclass[review=false]{acmart}
\usepackage{booktabs} 
\usepackage{amsfonts}       
\usepackage{amsmath}
\usepackage[ruled, vlined, linesnumbered]{algorithm2e}

\usepackage{graphicx}
\usepackage{mathrsfs}
\usepackage{enumerate}
\usepackage{multirow}
\usepackage{enumitem}
\usepackage{todonotes}
\usepackage{amsthm}
\usepackage{subcaption}
\usepackage{bbm}
\usepackage{pifont}
\usepackage{array}
\newcolumntype{P}[1]{>{\centering\arraybackslash}p{#1}}

\newcommand{\para}[1]{{\vspace{5pt} \bf \noindent #1 \hspace{8pt}}}

\theoremstyle{definition}
\newtheorem{defn}{Definition}

\setcopyright{none}
\acmYear{2017}
\copyrightyear{2017}

\begin{document}
\title{Crowdsourcing Cybersecurity: \\ Cyber Attack Detection 
using Social Media}

\author{Rupinder Paul Khandpur\textsuperscript{1,2}, Taoran Ji\textsuperscript{1,2}, Steve Jan\textsuperscript{3}, \\
Gang Wang\textsuperscript{3}, Chang-Tien Lu\textsuperscript{1,2}, Naren Ramakrishnan\textsuperscript{1,2}}
\affiliation{%
	\institution{\\ \textsuperscript{1}Discovery Analytics Center, Virginia Tech,
	Arlington, VA 22203, USA\\
	\textsuperscript{2}Department of Computer Science, Virginia Tech, Arlington, VA 22203, USA\\
	\textsuperscript{3}Department of Computer Science, Virginia Tech, Blacksburg, VA 24060, USA}
}
\renewcommand{\shortauthors}{R. Khandpur et.\ al.}

\begin{abstract}
Social media is often viewed as a sensor into various societal events
such as disease outbreaks, protests, and elections. We describe
the use of social media as a crowdsourced sensor to gain insight into
ongoing cyber-attacks. Our approach detects a broad range of
cyber-attacks (e.g., distributed denial of service (DDOS) attacks,
data breaches, and account hijacking) in an unsupervised manner using
just a limited fixed set of seed event triggers. A new query expansion
strategy based on convolutional kernels and dependency parses helps
model reporting structure and aids in identifying key event
characteristics. Through a large-scale analysis over Twitter,
we demonstrate that our
approach consistently identifies and encodes events, outperforming
existing methods.
\end{abstract}

\maketitle

\section{Introduction}

Today's widespread incidences of cyber-attacks (e.g., most recently of the
US Democratic Party and at companies such as 
Sony, Verizon, Yahoo, Target, JP Morgan, Office of Personnel Management, Ashley Madison), each more audacious than earlier ones, makes them perhaps the primary threat faced by individuals, organizations, and nations alike. Consequences and implications of cyber-attacks include monetary losses, threats to critical infrastructure and national security, disruptions to daily life, a potential to cause loss of life and physical property, and data leaks exposing sensitive personal information about users and their activities. The largely quasi-static and unadaptable nature of existing cyber-defenses makes them vulnerable to rapidly evolving attack mechanisms, and thus engender not just a `warning deficit' but also a 
`detection deficit', i.e., an increase in the mean time between the time of an attack and its discovery. 

It has been well argued that, because news about an organization's compromise 
sometimes originates {\it outside} the organization, one could use
open source indicators (e.g., news and social media) as indicators of
a cyber-attack. 
Social media, in particular, turns users into social sensors
empowering them to participate in an online ecosystem of event detection
for happenings such as 
disease outbreaks~\cite{signorini2011use},
civil unrest~\cite{zhao2014unsupervised, muthiah2015planned}, and
earthquakes~\cite{sakaki2010earthquake}. While the use of social media 
cannot fully supplant the need for internal telemetry for
certain types of attacks (e.g., use of network flow data to detect
malicious network behavior~\cite{Noble:2003:GAD:956750.956831, Kwon:2015:DEI:2810103.2813724,
Davis:2011:DAG:2063576.2063749}), analysis of such online media can
provide insight into a broader range of cyber-attacks such as data
breaches, account hijacking and newer ones as they emerge.

At the same time it is non-trivial to harness social media to identify 
cyber-attacks. {\bf Our objective is to detect a range of different
cyber-attacks as early as possible, determine their characteristics (e.g.,
the target, the type of attack), in an unsupervised manner.} Prior work
(e.g.,~\cite{Ritter:2015:WSE}) relies on weak supervision techniques which
will be unable to capture the dynamically evolving nature of cyber-attacks
over time and are also unable to encode characteristics of detected
events, as we aim to do here.

Our main contributions are:
\begin{itemize}
    \item \textbf{A framework for cybersecurity event detection based on
        online social media.} Our dynamic event trigger expansion (DETE)
approach uses a limited, fixed, set of general seed event triggers and
learns to map them to specific event-related expansions and thus provide
situational awareness into cyber-events in an unsupervised manner.
\item \textbf{A novel query expansion strategy based on dependency
tree patterns.}
To model typical reporting structure in how cyber-attacks are
described in social media,
        we propose a dynamic event trigger expansion method based on convolution
        kernels and dependency parses. The proposed
approach also employs a word embedding strategy to capture
similarities between event triggers and candidate event reports.
\item \textbf{Extensive empirical evaluation for three kinds of cyber-attacks}.
We manually catalog ground truth for three event 
classes---distributed denial of service (DDOS) attacks, data breaches, 
and account hijacking---and
demonstrate that our approach consistently identifies and encodes events
outperforming existing methods.
\end{itemize}

\section{Problem Setup}
\label{sec:problem_setup}
The input to our methodology is
a collection of time-ordered tweets
$\mathbb{D} = \{\mathbb{D}_1, \mathbb{D}_2, \ldots, \mathbb{D}_p\}$ organized
along $p$ time slots. Let $\mathcal{D}$ denote the tweet space corresponding
to a subcollection $\mathbb{D}_{i}$, let $\mathcal{D}^{+}$ denote the 
target tweet subspace (in our case,
comprising cyber-attack events), and let
$\mathcal{D}_{-} = \mathcal{D} - \mathcal{D}_{+}$ denote the
rest of the tweets in the considered tweet space. 

\begin{defn}
    \textbf{Typed Dependency Query}: A \textbf{\textit{typed dependency query}}
    is a linguistic structure that characterizes a semantically coherent event
    related topic. Different from n-grams, terms contained in a
    \textbf{\textit{typed dependency query}} share both syntactic and semantic
    relationships. Mathematically, a \textbf{\textit{typed dependency query}} is
    formulated as a tree structure $G = \{V, E\}$, where node $v \in V$ can be
    either a unigram, user mention, or a hashtag and $\varepsilon \in E$
    represents a syntactic relation between two nodes.
\end{defn}

\begin{defn}
    \textbf{Seed Query}: A \textbf{\textit{seed query}} is a manually selected
    typed dependency query targeted for a certain type of event. For instance,
    ``hacked account'' can be defined as a potential \textbf{\textit{seed query}} for
    an account hijacking event.
\end{defn}

\begin{defn}
    \textbf{Expanded Query}: An \textbf{\textit{expanded query}} is a typed
    dependency query which is automatically generated by the DETE algorithm based
    on a set of seed queries and a given
tweet collection $\mathcal{D}$. 
    \textbf{\textit{expanded query}} and its seed query can be two different
    descriptions of the same subject. More commonly,
    an \textbf{\textit{expanded query}} can be more specific than its seed query.
    For instance, ``prime minister dmitry medvedev twitter account hack'', an
    expanded query from ``hacked account'', denotes the message of an account
    hijacking event related with Dmitry Medvedev.
\end{defn}

\begin{defn}
    \textbf{Event Representation}: An event $e$ is defined as
    $(\mathcal{Q}_{e}, date, type)$, where $\mathcal{Q}_{e}$ is the set of
    event-related expanded queries, $date$ denotes when the event happens,
    and $type$ refers to the category of the cyber-attack event
(i.e., DDOS, account hijacking, or data breach).
\end{defn}

Here $\mathcal{Q}_{e}$ is a defined as a set because, in general, a cyber-attack
event can be presented and retrieved by multiple query templates. For instance, 
among online discussion and report about event ``Fashola's account, website
hacked'', the query template most used are ``fashola twitter account hack'',
``fashola n78m website twitter account hack'' and ``hack account''.

Given the above definitions, the major tasks underlying the cyber-attack event detection
problem are defined as follows:

\textbf{Task 1:} \textbf{Target Domain Generation:} Given a tweet
subcollection $\mathcal{D}$, \textbf{\textit{target domain generation}} is the
task of identifying the set of target related tweets $\mathcal{D}_{+}$. $\mathcal{D}_{+}$
contains critical target related information based on which the expanded query can
be mined.

\textbf{Task 2:} \textbf{Expanded Query Extraction:} Given target domain
$\mathcal{D}_{+}$, the task of 
\textbf{\textit{expanded query extraction}} is to generate a
set of expanded queries $\mathcal{Q} = \{q_{1}, \ldots, q_{n}\}$
which represents the generic concept delivered by $\mathcal{D}_{+}$. Thus set
$\mathcal{Q}$ can be used to retrieve event related information from other
collection sets.

\textbf{Task 3:} \textbf{Dynamic Typed Query Expansion:} Given a small set
of seed queries $\mathcal{Q}^{0}$ and a twitter collection $\mathcal{D}$,
the task of
\textbf{\textit{dynamic typed query expansion}} is to iteratively
expand $\mathcal{D}_{+}^{k}$ and $\mathcal{Q}^{k}$ until all the target
related messages are included.

\section{Methodology}

In traditional information extraction (IE), a large corpus of text must first be
annotated to train extractors for event triggers, defined as main
keywords indicating an event occurrence~\cite{ji2008refining}. However, in our
scenario using online social media, a manually annotated label set is impractical
due to the huge volume of online media and the generally
noisy characteristics of the text. In this
section, we propose a novel method to automatically mine query templates over
which the event tracking is performed.

\subsection{Target Domain Generation}
\label{sub:target_domain_generation}

In this subsection, we propose the method of target domain generation, which
serves as the source of social indicators for the detection of ongoing
cyber-attack events. Given a query and a collection of tweets $\mathcal{D}$, the
typical way to retrieve query-related documentation is based on a bag of words
model~\cite{salton1986introduction} which comes with its attendant
disadvantages.
Consider the following two
tweets: ``has riseups servers been compromised or \textbf{data leaked}?'' and
``@O2 You completely screwed me over! My phones back on, still \textbf{leaking
data} and YOU are so UNHELPFUL \#CancellingContract \#Bye''. Though the
important indicator ``leak data'' for data breach attack is involved in both
tweets, the second tweet is complaining
about a phone carrier and is unexpected noise
in our case. To address this problem, syntactically bound information and 
semantic similarity constraints are jointly considered in our proposed method.

More specifically, each tweet in $\mathcal{D}$ is first
converted into its dependency tree form. Thus for a given seed query $q$, 
the target domain $\mathcal{D}_{+} \subseteq \mathcal{D}$ can be generated by collecting
all tweets which are both syntactically and semantically similar to the seed query $q$.
Mathematically, given two dependency trees $q$ and $d$, a
convolution tree 
kernel~\cite{kate2008dependency} is adopted to measure the similarity using
shared longest common paths:
\begin{equation}\label{equ:tree_kernel}
    \mathrm{K}(q, d) = \underset{\substack{u \in q \\ v \in d}}\sum
    {\big(1 + \mathcal{H}(u, v)\big)}^{\mathbbm{1}_{\mathbb{R}_{>0}}\big(
    \mathcal{H}(u, v)\big)}
\end{equation}
where $v$ and $u$ are two nodes from two trees $q$ and $d$ respectively,
$\mathbb{R}_{>0}$ represents set of positive real numbers, $\mathbbm{1}(\cdot)$
is the indicator function and $\mathcal{H}(v,u)$ counts the number of common
paths between the two trees which peak at $v$ and $u$, which can be calculated
by an efficient algorithm proposed by Kate et al.~\cite{kate2008dependency}, as
described in Algorithm~\ref{alg:CPP}.

\begin{algorithm}[htbp]\small
\SetAlgoSkip{}
\SetAlgoVlined%
\LinesNumbered%
\KwIn{$u \in q, v \in d$}
\KwOut{$\mathcal{H}(u, v)$}
Set $count = 0, r = \kappa(u, v)$\;
Set $\mathcal{C}_{u} = children(u)$\;
Set $\mathcal{C}_{v} = children(v)$\;
\For{$c_{i}, c_{j} \in \mathcal{C}_{u}, i \neq j$}
{%
    \For{$c_{m}, c_{n} \in \mathcal{C}_{v}, m \neq n$}
    {%
        \If{$c_{i} \doteq c_{m}$ and $c_{j} \doteq c_{n}$}
        {%
            $x = \kappa(c_{i}, c_{m})$\;
            $y = \kappa(c_{j}, c_{n})$\;
            $r = r + \sqrt{\lambda} + \lambda x + \lambda y + \lambda xy$
        }
    }
}
$\mathcal{H}(u, v) = r$\;
\caption{Calculation of number of common paths.}\label{alg:CPP}
\end{algorithm}

In Algorithm~\ref{alg:CPP}, $\kappa(u, v)$ is the number of common paths
between the two trees which originate from $u$ and $v$, and can be recursively
defined as:
\begin{equation}
\label{equ:CDP}
    \kappa(u, v) = \underset{\substack{\mu \in C(u) \\ \eta \in C(v)}}
    \sum{(1 + \kappa(\mu, \eta))}^{{\mathbbm{1}_{\mu \doteq \eta}(u, v)}},
\end{equation}
where $C(\cdot)$ denotes the set of children node. In both
Algorithm~\ref{alg:CPP} and Equation~\ref{equ:CDP}, we use the semantic
similarity operator $\doteq$, introduced to consider the semantic similarity of
tree structre. This semantic similarity is computed by considering cosine similarity of word embeddings
vector generated from the word2vec algorithm.

This model considers the common paths which are linguistically meaningful, which
reduces the noise introduced by coincidentally matched word chains. In addition,
long-range dependencies between words, which decreases the performance, are
avoided because functionally related words are always directly linked in a 
dependency tree. 

\subsection{Dynamic Typed Query Expansion}
\label{sub:dynamic_typed_query_expansion}

In this subsection, we propose a way to dynamically mine an
expanded query given a
small collect of seed query, as shown in Table~\ref{tab:seedquery}. By
providing a small set of seed queries (unigrams), 
Zhao et al.~\cite{zhao2014unsupervised} proposed a dynamic
query expansion (DQE) method which is able to iteratively expand the seed query
set from currently selected target tweet subspace until
convergence.
Looking beyond the simple unigrams based
expansion, by introducing dependency-based tree structure extraction, we build a
dynamic expanded query generation model for the
cyber-attack detection task. 

\begin{table}[htpb]
    \centering
    \caption{Seed queries for cyber-attack events.}
    \begin{tabular}{cp{160pt}} 
        \toprule
        \textbf{Category} & \multicolumn{1}{c}{\textbf{Seed Query}}\\
        \midrule                                                           
        Data breach & data leak, security breach, information stolen,
        password stolen, hacker stole\\ 
        \midrule
        DDoS &  DDoS attack, slow internet, network infiltrated, malicious
        activity, vulnerability exploit, phishing attack\\
        \midrule      
        Account Hijacking & unauthorized access, stolen identity, hacked
        account\\
        \bottomrule           
    \end{tabular}\label{tab:seedquery}
\end{table}

Let us denote $\mathcal{Q}^{k}$, $\mathcal{D}_{+}^{k}$ as the
expanded query set and target domain at the $k$th iteration. Before the
iteration process, $\mathcal{Q}^{0}$ is initialized by the manually selected
small set of seed queries, as shown in Table~\ref{tab:seedquery}. With
$\mathcal{D}$ and $\mathcal{Q}^{0}$, then $\mathcal{D}_{+}^{0}$, the target
domain at iteration $0$, is generated based on the convolution tree kernel, as
described in Equation~\ref{equ:tree_kernel}. At the $k$th iteration, given
last expanded query set $\mathcal{Q}^{k - 1}$ and last generated target domain 
$\mathcal{D}_{+}^{k - 1}$, our approach first prepares candidate expanded
queries for each matched $q_{i} \in \mathcal{Q}^{k - 1}$ and
$d \in \mathcal{D}_{+}^{k - 1}$:
\begin{equation}
    \hat{q}_{i}^{k} = \mathrm{subgraph}\big(
    \underset{v \in d}{\text{argmax}}(\sum_{u \in q_{i}}\mathcal{H}(v, u))\big),  
\end{equation}
where $v$ and $u$ are term nodes in tweet $d$ and $q_{i}$ respectively, and
$\mathrm{subgraph}(\cdot)$ is an operator to extract the subtree structure from
entire tree with $v$ as root. Thus the candidate query expansions are collected
based on relevant document and query space, that is
$\mathcal{D}_{+}^{k - 1}$ and $\mathcal{Q}^{k - 1}$. To identify the best
candidate expanded queries, query terms are then ranked based on
Kullback-Leibler distance~\cite{mei2005discovering} between target domain
$\mathcal{D}_{+}^{k-1}$ and the whole tweet collection $\mathcal{D}$:
\begin{equation}
    KL(f, \mathcal{D}_{+}^{k - 1}|\mathcal{D}) =
    \log\frac{\Pr(f | \mathcal{D}_{+}^{k-1})}{\Pr(f | \mathcal{D})}
    \Pr(f | \mathcal{D}_{+}^{k - 1}),
\end{equation}
where $KL(f, \mathcal{D}_{+}^{k - 1}|\mathcal{D})$ denotes the Kullback-Leibler
distance, $f$ is any term in $\hat{q}_{i}^{k}$, $\Pr(f|\mathcal{D}_{+}^{k-1})$
and $\Pr(f|\mathcal{D})$ is the probability of term $f$ appears in 
$\mathcal{D}_{+}^{k - 1}$ and $\mathcal{D}$ respectively. Using the $KL$
distance to rank query terms we are able to assign scores to terms that best
discriminate relevant and non-relevant expansions. For example query terms such
``account'' and ``twitter'' both appear frequently in the candidate expansions
but they have little informative value as they will have the same (random)
distribution in any subset of the twitter collection, whereas terms such as
``hacked'' will have comparatively higher probability of occurence in the
relevant subspace. These high ranked candidates will then act as the expanded
queries set to run the next iteration until the algorithm converges. The
detailed dynamic typed query expansion algorithm is shown in
Algorithm~\ref{alg:DQE}.

\begin{algorithm}[!htbp]\small
\SetAlgoSkip{}
\SetAlgoVlined%
\LinesNumbered%
\KwIn{Seed query $\mathcal{Q}^{0}$, Twitter subcollection $\mathcal{D}$}
\KwOut{Expanded Query $\mathcal{Q}$}
Set $D_{+}^{0} = match(\mathcal{Q}^{0}, \mathcal{D}), k = 0$\;
\Repeat{$\bigcup\limits_{i=0}^{k} \mathcal{Q}^{i} - \bigcup\limits_{i=0}^{k-1} \mathcal{Q}^{i} \neq \emptyset$\tcp*[h]{DQE iteration}}
{%
    $k = k + 1$\;
    \For{$q_{i} \in \mathcal{Q}^{k - 1}$}
    {%
        $\widehat{q}^{k}_{i} = subtree(\underset{v \in d}
        {\mathrm{argmax}}\underset{u \in q_{i}}\sum CPP(v, u))$\;
        \For{$f \in \widehat{q}^{k}_{i}$} 
        {%
            $\Pr(f | \mathcal{D}_{+}^{k - 1}) = \frac{tf(f)}{
                |\mathcal{D}_{+}^{k-1}|}$\;
            $\Pr(f | \mathcal{D}) = \frac{tf(f)}{|\mathcal{D}|}$\;
            $w(f) = KL(f, \mathcal{D}_{+}^{k-1} | \mathcal{D})$\;
        }
        $w(\widehat{q}^{k}_{i}) = \sum_{f \in \widehat{q}^{k}_{i}}w(f)$\;
    }
    $\mathcal{Q}^{k} = topK(w(\widehat{\mathcal{Q}}^{k})),
    \widehat{\mathcal{Q}}_{k} = \{\widehat{q}_{1}^{k}, \ldots,
    \widehat{q}_{|\widehat{\mathcal{Q}}^{k}|}^{k}\}$\;
    $\mathcal{D}^{k} = match(\mathcal{Q}^{k}, \mathcal{D})$\;
}
$\mathcal{Q} = \mathcal{Q}^{k}$\;
\caption{Dynamic Typed Query Expansion Algorithm.}\label{alg:DQE}
\end{algorithm}

\subsection{Event Extraction}
Given an expanded query set $\mathcal{Q}$, we extract $\mathcal{Q}_{s} \mid q_{i} \nsubseteq q_{j} \mid q_{i}, q_{j}\in \mathcal{Q}_{s}$. 
For example, consider the surface string representations of a set of expanded queries $Q$ as (``data breach'', ``data leak'', ``ashley madison'', ``ashley madison data breach'') then $ \mathcal{Q}_{s}$ will be (``ashley madison data breach'').
We then cluster the query expansions in $\mathcal{Q}_{s}$ using affinity propagation~\cite{frey2007clustering} and also extract \textit{exemplars} $q_{e}$, of each query set $\mathcal{Q}_{e}$ that are representative of clusters, where each member query is represented by a vector $\boldsymbol{\tilde{q}}$ calculated from the word embedding $\boldsymbol{\tilde{u}}$ of the lemma of each query term $u \in q$ as:
\begin{equation}
    \boldsymbol{\tilde{q}} = \sum_{u \in q}\boldsymbol{\tilde{u}}.
\end{equation}

Each exemplar query $q_{e}$ is then annotated to a cyber-attack type. For 
this purpose, we first compute the cosine similarity between an exemplar query expansion $q_{e}$ and seed
query $q_{j} \in \mathcal{Q}^{(0)}$ as:
\begin{equation}
    \mathrm{sim}(q_{e}, q_{j}) = \frac{%
        \boldsymbol{\tilde{q_{e}}}\cdot \boldsymbol{\tilde{q_{j}}}}
    {||\boldsymbol{\tilde{q_{e}}}||\cdot||\boldsymbol{\tilde{q_{j}}}||}.   
\end{equation}
The $q_{j} \in \mathcal{Q}^{(0)}$ which has the highest similarity value
with $q_{e}$ determines the event type to which $\mathcal{Q}_{e}$ belongs to. For the complete event representation $(\mathcal{Q}_{e}, date, type)$ date information is extracted based on the time interval choosen for DQE; for example in
our experiments we run DQE on a daily aggregated collection of tweets. In this way we extract the final set of event tuples.

\begin{figure*}[!htpb]
\centering
\hspace*{\fill}
\includegraphics[width=\linewidth]{%
    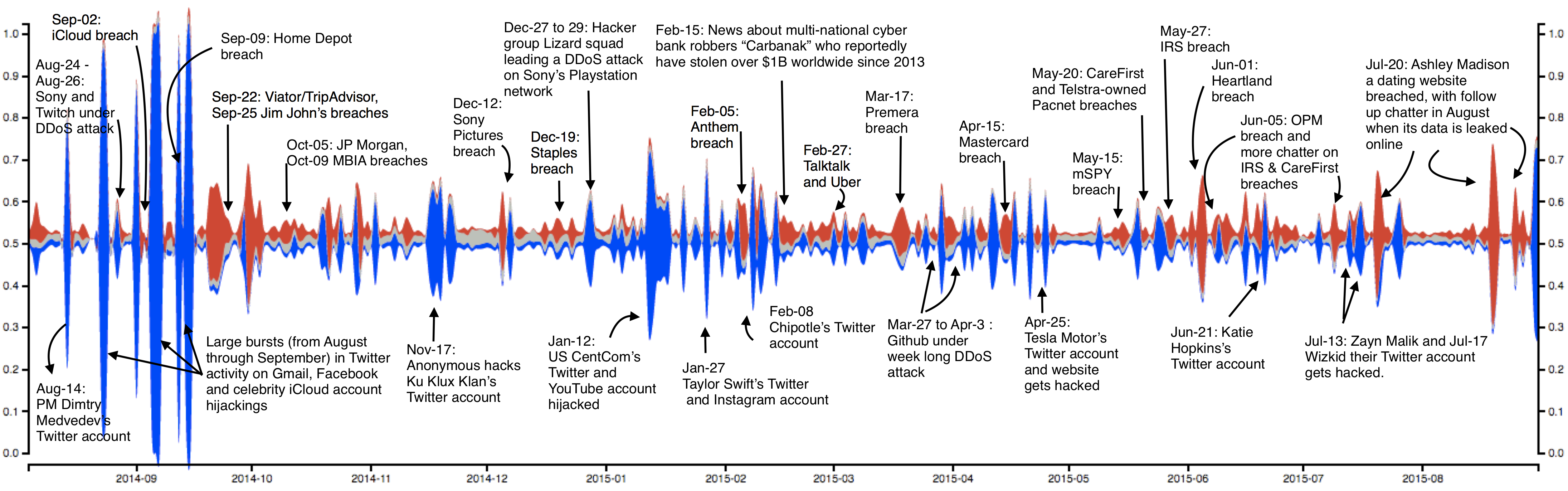}
\caption{Streamgraph showing normalized volume of tweets (August, 2014 through
    August, 2015) tagged with data breach (red), DDoS activity (grey) and
    account hijacking (blue) type of cyber-security events.}
\label{fig:dqestreamgraph2014}
\end{figure*}

\begin{figure*}[!htpb]
\centering
\hspace*{\fill}
\includegraphics[width=\linewidth]{%
    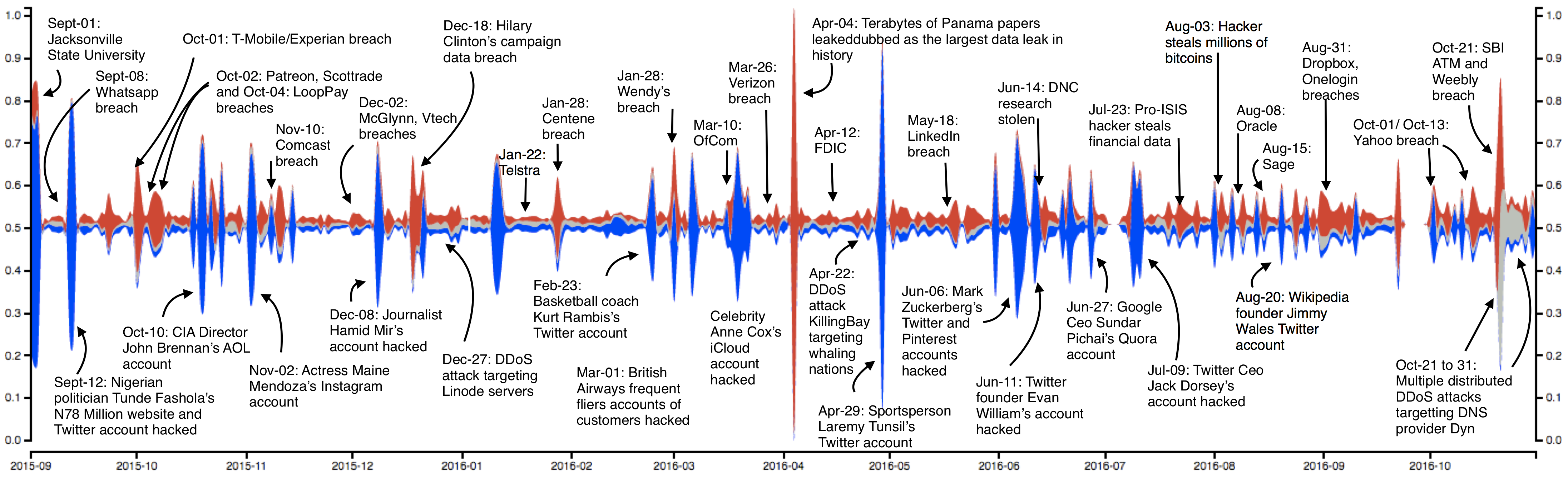}
\caption{Streamgraph showing normalized volume of tweets (September, 2015
    through October, 2016) tagged with data breach (red), DDoS activity (grey)
    and account hijacking (blue) type of cyber-security events.}
\label{fig:dqestreamgraph2016}
\end{figure*}

\section{Evaluation}

\subsection{Evaluation Setup}
\label{sub:evaluation_setup}

\subsubsection{Dataset and Gold Standard Report}
\label{ssub:dataset_and_gold_standard_report}

We evaluate the proposed method on a large stream of tweets from GNIP's decahose
(10\% sample) collected from August, 2014 through October, 2016. The
total raw volume of our Twitter dataset across these 27 months is
5,146,666,178 (after removing all retweets). Then, from this raw volume we create 2 subset collections:  
\begin{itemize}
    \item {\bf Fixed Keyword Filtered Tweets:}  We filtered 79,501,789 tweets
      that contain at least one matching term from a list
      of cyber attack related keywords. These are top 1000 keywords (ranked by TFIDF) extracted from
      description texts of events in our gold standard report
      (see below).
    \item {\bf Normalized Tweet Texts:} We extract and normalize tweet
      texts (after removing accents, user mentions and urls) to
      produce a collection of 3,267,567,087 unique texts to train a
      200 dimensional word embedding via gensim's word2vec
      software~\cite{rehurek2010software}.
\end{itemize}
Note that the experimental results for the performance of our typed
dynamic query expansion algorithm are done using the entire raw volume
of over 5 billion tweets.
The total volume of
tweets filtered from query expansion algorithm is 1,093,716 over
the entire time period. 

To evaluate our methods, we need Gold Standard Reports (GSR) on cyber security incidents to serve as ground-truth. In particular,
we focus on highly impactful events on data breach, DDoS and account hijacking based on two
different sources, Hackmageddon~\cite{WinNT} and
PrivacyRights~\cite{PrivacryRight}. In both sources, each event is characterized by an event type, date (when the event was publicly
    reported), victim organizations and a short description. 
\begin{itemize}
\item \textbf{Hackmageddon} is an independently maintained
  website that collects public reports of cyber 
    security incidents. Between January 2014 and
    December 2016, we extract 295 account
    hijacking events and 268 DDoS events. 
For account hijacking, since we are using social media data, we mainly focus on hijacking attacks on social media accounts (twitter,
    instagram, facebook) by cyber crimes. After filering US-based
    events and matching
    the time range of our Twitter data, we obtain 55 account hijacking
    events and 80 DDoS events for GSR. 
    \item \textbf{PrivacyRights} is a highly reputable repository for data breach 
    incident reports. 
Between January 2014 and December 2016, we extract 1064 data breach events reported by various sources. To enhance the
    accuracy of GSR, we choose events reported by four large, well-known sources ---
    ``Media'', ``KrebsOnSecurity'', ``California Attorney General'', and
    ``Security Breach Letter''. Then, we filter out data
    breaches caused non-cyber reasons ({\em e.g.}, physical theft) and
    focus on the HACK category. After removing events with an unknown
    size of data loss, and matching the time range
    with our data, we have 85 data breach events for GSR. 
\end{itemize}

\subsubsection{Baseline and Comparison Method}
\label{ssub:baseline_and_comparison_method}

Our baseline for event detection uses the well-known Kleinberg's
algorithm~\cite{Kleinberg:2002:BHS:775047.775061}. It identifies time periods in
which a target event is uncharacteristically frequent or ``bursty'' on a set of
static keywords. We extract an event if the size of
this set of ``bursty'' keywords is larger than a threshold $T_b$. 

In this experiment, we use the 79.5 million Fixed Keyword Filtered
Tweets and the 1000 static keywords to run the baseline method. We set
the threshold $T_b$ based on small scale empirical tests on a few
months of data, and manually examine the detected events. We set $T_b=$36 which
returns a better event/noise ratio. We apply this threshold on all the
data and detects 81 events from August, 2014 through October, 2016. Each detected event is characterized as by
a date and a set of ``bursty'' keywords.

\subsubsection{Matching Detected Events with GSR}
\label{ssub:baseline_and_comparison_method}

Given a detected event presented by $e = (\mathcal{Q}_{e}, date, type)$, we
developed a semi-automatic method to detect if $e$ is matchched with any event
in GSR:\ 
\begin{enumerate}
    \item For named entity in $e$, we check if it matches any event description
        in GSR, and get a matched collection from GSR, say $ME$,
    \item Further filter $ME$ by matching the event date between
        $date$ in $e$ and $ME$, with a time window as 3 (one day
        before $date$, and one day after $date$), and get a new filted event
        set, say $FME$,
    \item Compare the event type between $e$ and event in $FME$, if the event
        type also matches, then event $e$ is consider as a matched event.
\end{enumerate}

However, considering that the detected events are mined from Twitter
environment which may not use formal keywords to describe the event. We will manually
double check the event $e$ if it fails the step 1. 

Detected events by baseline will follow the same method to match against
GSR. The only adjustment is to match the ``bursty'' keywords of the
detected events instead of named entities.

\subsection{Measuring Performance}
\label{sub:measuring_performance}

Precision and recall over different types of cyberattack events are summarized
in Table~\ref{tab:precision_recall}. These results shows that with only a small
set of seed query templates (as shown in Table~\ref{tab:seedquery}), our
approach can reach around $80\%$ of precision for data breach and
DDoS events. This means our approach is able to handle the noisy Twitter environment and
perform the cyberattack event detection accurately. The precision for
account hijacking is not as high ($66\%$). Careful manual analysis
(through Google search) indicates that we actually detected new account hijacking events that are not covered by the GSR
(Table~\ref{tab:tp_fp_ah}). The manual validation results for data
breach and DDoS events are shown in Table~\ref{tab:tp_fp_db} and
Table~\ref{tab:tp_fp_ddos}. We also detected new events that are not
covered by GSR for these two types.  

Data breach events have a higher recall ($75\%$). The relatively low
recall for account hijacking and DDoS is explainable. 
Both DDoS and account hijacking events have a rather short life cycle from occurring 
to being addressed. Thus their signal in social media is relatively
weaker. For instance, DDoS often cause a few minutes to several hours of
slow internet, which may end even before people realized it. 
This intuition is validated in the baseline performance. The extremely low precision and
recall in Table~\ref{tab:precision_recall_baseline} shows relying on
``burstiness'' is difficult to capture such events, possibly due to
their weak signals over noise.

\begin{table}[htpb]
    \centering
    \begin{tabular}{lcc}
        \toprule  
        Category & Number of Detected Events & GSR\\
        \midrule
        Data Breach &  \textbf{227} & 85\\
        Account Hijacking & \textbf{127} &55\\
        DDoS Attacks & \textbf{109} & 80 \\
        Duplicate & 606 & 0\\
        Rejected & 581 & 0\\
        Unspecified & 390 & 0\\
        \midrule
        Total & 2040 & 220\\
        \bottomrule
    \end{tabular}
    \caption{Number of cyber attack events in GSR and those detected by typed DQE.}\label{tab:events_statistic}
\end{table}

\begin{table}[htpb]
    \centering
    \begin{tabular}{ccc}
        \toprule  
        \multirow{2}{*}{Manually Verified} & 
        \multicolumn{2}{c}{Matched with GSR} \\
        \cmidrule{2-3}
        & Yes & No\\
        \midrule
        True Positive  &  22 & 156\\
        False Positive &  0  & 49\\
        \bottomrule
    \end{tabular}
    \caption{Events detected of data breach type.}\label{tab:tp_fp_db}
\end{table}

\begin{table}[htpb]
    \centering
    \begin{tabular}{ccc}
        \toprule  
        \multirow{2}{*}{Manually Verified} & 
        \multicolumn{2}{c}{Matched with GSR} \\
        \cmidrule{2-3}
        & Yes & No\\
        \midrule
        True Positive  &  8 & 51\\
        False Positive &  0  & 31\\
        \bottomrule
    \end{tabular}
    \caption{Events detected of account hijacking type.}\label{tab:tp_fp_ah}
\end{table}

\begin{table}[htpb]
    \centering
    \begin{tabular}{ccc}
        \toprule  
        \multirow{2}{*}{Manually Verified} & 
        \multicolumn{2}{c}{Matched with GSR} \\
        \cmidrule{2-3}
        & Yes & No\\
        \midrule
        True Positive  &  20 & 29 \\
        False Positive &  0  & 12 \\
        \bottomrule
    \end{tabular}
    \caption{Events detected of DDoS attack type.}\label{tab:tp_fp_ddos}
\end{table}

\begin{table}[htpb]
    \centering
    \begin{tabular}{cccc}
        \toprule  
        Event Type & Precision & Recall & F-measure \\
        \midrule
        Data Breach         &  0.78 & 0.74 & 0.76\\
        Account Hijacking   &  0.66 & 0.56 & 0.64\\
        DDoS Attack   &  0.80 & 0.45 & 0.58\\
        \bottomrule
    \end{tabular}
    \caption{Typed DQE based event detection performance comparison.}\label{tab:precision_recall}
\end{table}
\begin{table}[htpb]
    \centering
    \begin{tabular}{cccc}
        \toprule  
        Event Type & Precision & Recall & F-measure \\
        \midrule
        Data Breach         &  0.21 & 0.20 & 0.20\\
        Account Hijacking   &  0.01 & 0.02 & 0.01\\
        DDoS   &  0.01 & 0.01 & 0.01\\
        \bottomrule
    \end{tabular}
    \caption{Baseline's event detection performance comparison.}\label{tab:precision_recall_baseline}
\end{table}

\begin{figure*}[htpb]
    \begin{subfigure}[t]{0.40\textwidth}
        \includegraphics[height=1.6in, keepaspectratio]{%
            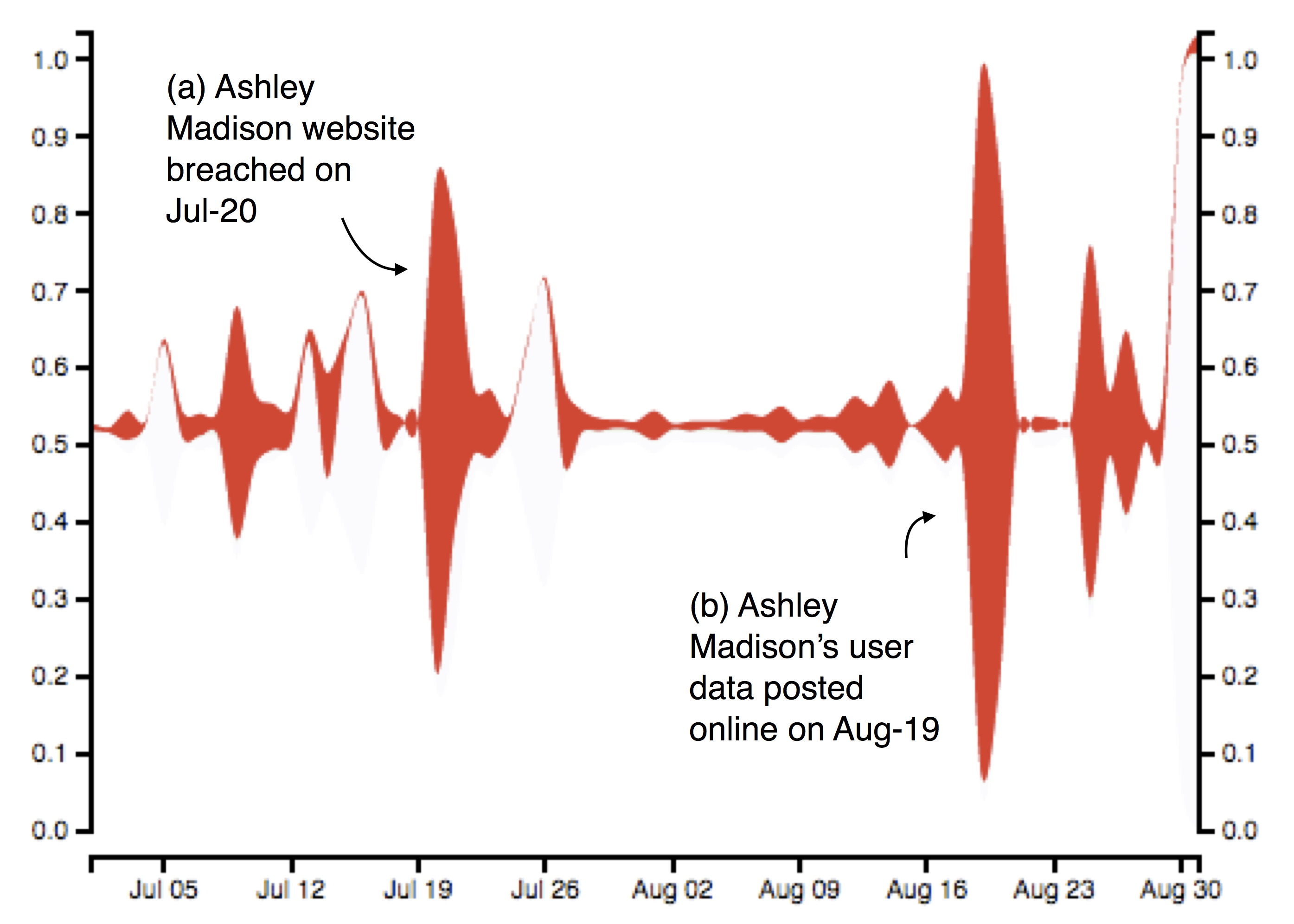}
    \end{subfigure}
    \begin{subfigure}[t]{0.58\textwidth}
        \includegraphics[height=1.7in]{%
            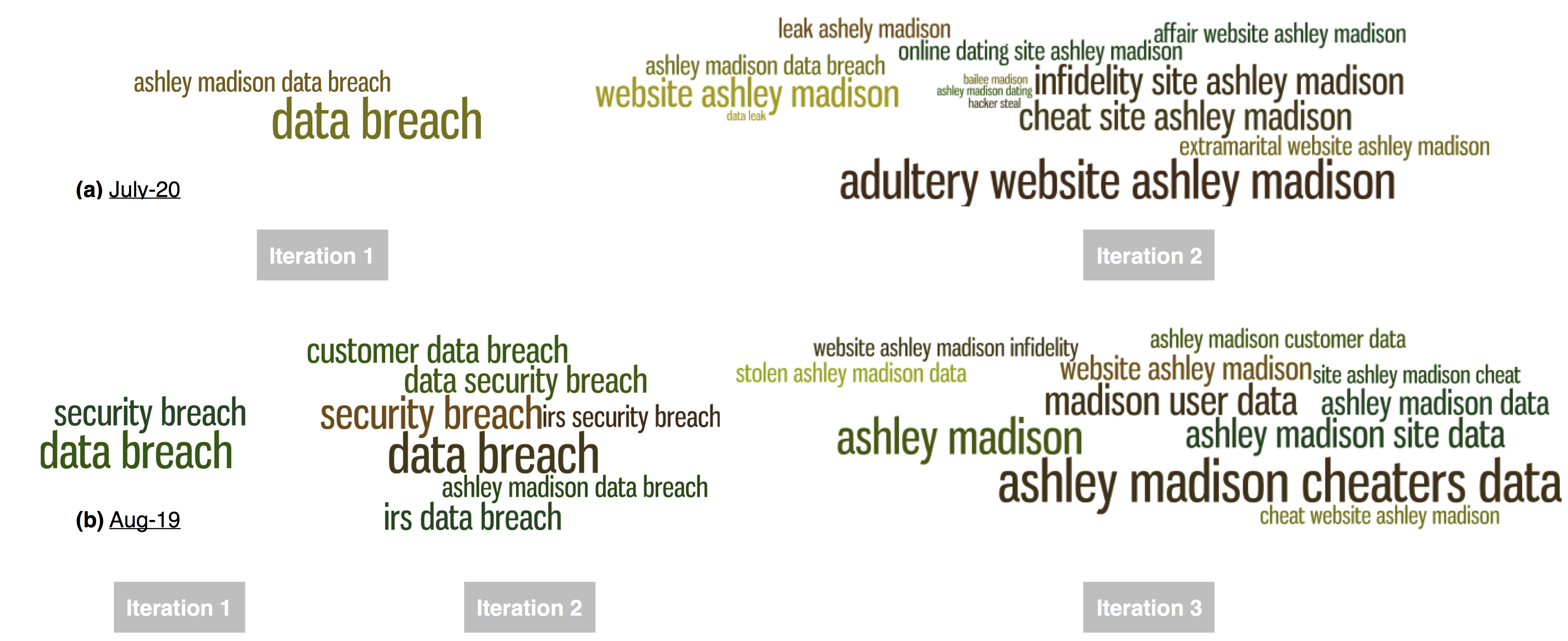}
    \end{subfigure}
    \caption{Word cloud and streamgraph analysis. The word cloud of all query
    expansions (size is proportional to the query's feature score) is produced
    from event detected for Ashley Madison website data breach. Streamgraph shows
    the bursty normalized volume of tweets when events happen.}\label{fig:ashleymadison}
\end{figure*}

\begin{figure}[htpb]
    \centering
    \includegraphics[height=1.2in,keepaspectratio]{%
        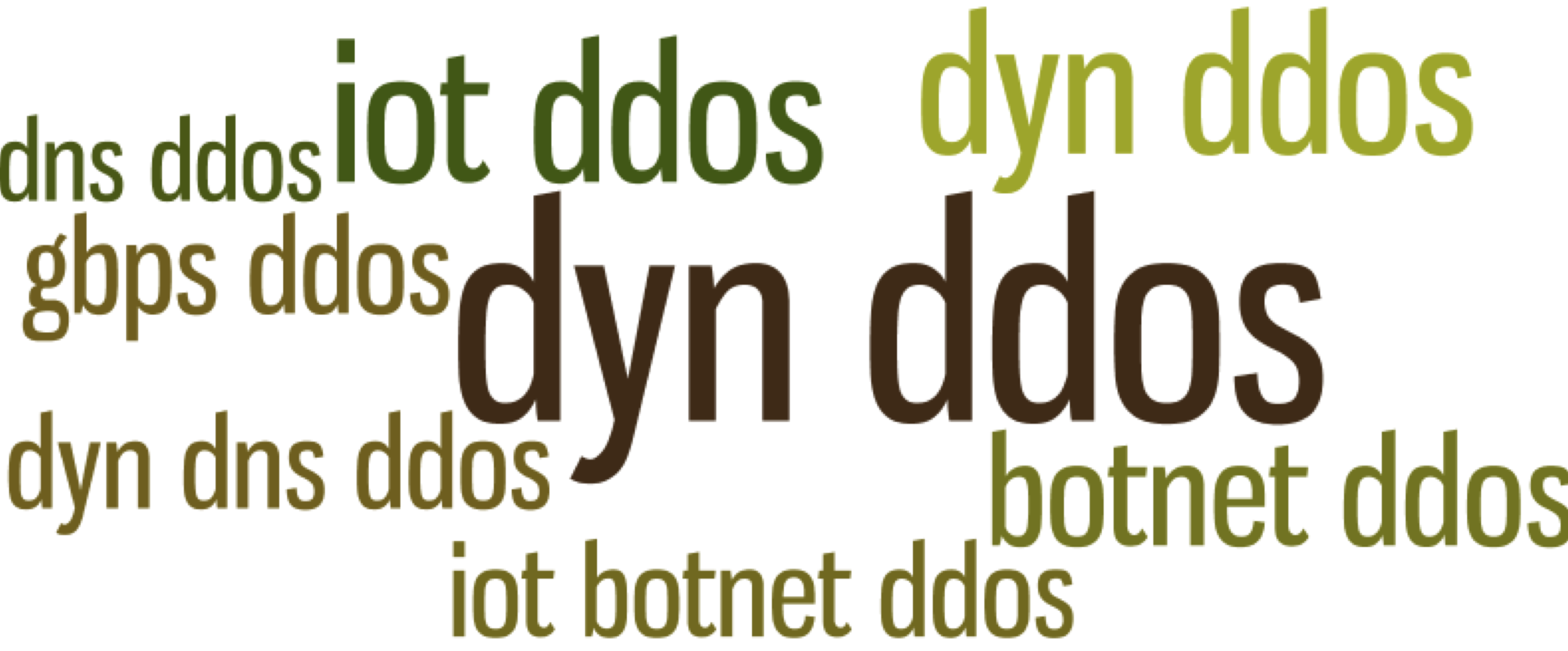}      
        \caption{A word cloud of all query expansions (size is proportional to the query's feature score) produced on  October 22, 2016 characterizing the DDoS attack on DNS provider Dyn.}\label{fig:dyndns}
\end{figure}

\subsection{Case Studies}
We comprehensively show in Fig.~\ref{fig:dqestreamgraph2014} and Fig.~\ref{fig:dqestreamgraph2016}, the wide range of events that our system is able to detect. Notice, the clear burst in Twitter activity that our query expansion algorithm is able to pick. Through the following case studies we highlight some of the interesting cases for each of three cyber attack types, that our system detected.

\textbf{Targeted DDoS Attacks on Sony and Dyn}: 
In late, November 2014, a hacker group calling itself ``The Guardians Of Peace'' hacked their way into Sony Pictures, leaving the Sony network crippled for days, allegedly perpertrated by North Korea. We capture 12 separate events of DDoS attacks including four in last week of August 2014, starting with the first on August 24th. Further in 2015, more ensuing attacks are captured one highlighted by the data breach of their movie production house, on December 12th and then a massively crippling targeted, DDoS attack on their PlayStation network in late December, 2015. Another noteworthy case of DDoS attacks in 2016, is the multiple distributed denial-of-service attack on DNS provider ``Dyn'' from October 21st through 31st in 2016, that almost caused an worldwide internet outage. Our system detects generates several query expansions, shown in Fig.~\ref{fig:dyndns} which clearly characterizes the nature of these DDoS attacks where the hackers turned a large number of internet-connected devices around the world in to botnets executing a distributed attack.

\textbf{Ashley Madison Website Data Breach:} In July 2015, a group calling
itself ``The Impact Team'' stole the user data of Ashley Madison, 
an adult dating website billed as enabling extramarital affairs. 
The hackers stole the website\'s all customer data and threatened to release the personally identifying information if the site was not immediately shut down. On 18 and 20 August, the group leaked more than 25 gigabytes of company data, including user details. We are able to detect this data breach on July 20, 2015. The word clouds in Fig.~\ref{fig:ashleymadison} 
clearly show how our method iteratively expands from the seed queries to the expanded queries in the last, iteration 3 capturing a very rich semantic aspect of the breach. After the initial burst as seen in the figure, we also see a second corresponding burst a month later, on August 20 when the user data is released anot now the top query expansion captured characterized by the mentions of user data leak of the same website.

\begin{figure}[htpb]
    \centering
    \includegraphics[height=1.2in]{%
        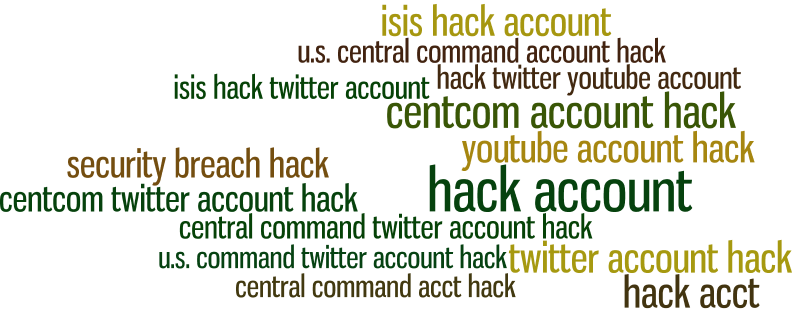}      
        \caption{A word cloud of all query expansions (size is proportional to the query's feature score) produced from event detected for U.S CentCom Twitter account hijacking.}\label{fig:uscentcom}
\end{figure}

\textbf{Twitter Account Hijackings}: We were also able to detect with very high date accuracy, several high profile cases of account hijackings of social media accounts of known personalities and government institutions including the Twitter account for U.S. Central Command which was hacked by ISIS sympathizers on January 12, 2015. We show in Fig.~\ref{fig:uscentcom} that our method not only identifies the victim (``central command twitter account hack'') but also the actor who perpetrated the hacking (``isis hack twitter account'').
\label{sub:case_studies}

\section{Related Work}

\para{Cyberattack Detection and Characterization.}
Detecting and characterizing cyber attacks is highly challenging due
to the constant-involving nature of cyber
criminals. Recent proposals cover a large range of different methods,
and Table~\ref{tab:related_work} lists representative works in
this space. Earlier work primarily focuses on mining network traffic data for intrusion detection. 
Specific techniques range from classifying malicious network
flows~\cite{Lee:1998:DMA} to anomaly detection in graphs to detect malicious
servers and connections~\cite{Ding:2012:ISC:2339530.2339670,Noble:2003:GAD:956750.956831,Kwon:2015:DEI:2810103.2813724,
  Davis:2011:DAG:2063576.2063749}. More recently, researchers seek to move ahead to predict cyber attacks
before they happened for early notifications~\cite{197270}. For example, Liu et
al.\ leverage various network data associated to an organization to look for
indicators of attacks~\cite{191002, Liu:2015:PCS:2713579.2713582}.  By
extracting signals from mis-configured DNS and BGP networks as well as spam and
phishing activities, they build classifiers to predict if an organization is (or
will be) under attack. Similarly, Soska et al.\ apply supervised classifiers to
network traffic data to detect vulnerable websites, and predict their chances of
turning malicious in the future~\cite{184495}.

In recent years, online media such as blogs and social networks become another
promising data source of security intelligence~\cite{7809731,aaai17}. Most
existing work focuses on technology blogs and tweets from
{\em security professionals\/} to extract useful information~\cite{Tsai2007}.
For example, Liao et al.\ builds text mining tools to extract key attack
identifiers (IP, MD5 hashes) from security tech
blogs~\cite{Liao:2016:AIG:2976749.2978315}. Sabottke et al.\ leverage Twitter
data to estimate the level of interest in existing CVE vulnerabilities, and
predict their chance of being exploited in practice~\cite{191006}. Our work 
differs from existing literature since we focus on crowdsourced data from the
much broader user populations who are likely the {\em victims\/} of security
attacks. The most related work to ours is~\cite{Ritter:2015:WSE} which uses
weakly supervised learning to detect security related tweets. However, this technique is unable to capture the dynamically
evolving nature of attacks and is unable to
encode characteristics of detected events.

\para{Event Extraction and Forecasting on Twitter.}
Another body of related work focuses on Twitter to extract various
events such as trending
news~\cite{Atefeh:2015:STE:2768330.2768335, 37918732918}, natural
disasters~\cite{Sakaki:2010}, criminal incidents~\cite{Wang2012} and population
migrations~\cite{6657174}. Common event extraction methods include simple
keyword matching and clustering, and topic modeling with temporal and
geolocation constrains~\cite{icwsm14event,icwsm11event,Zhou2014}. 
Event forecasting, on the other hand, aims to predict future evens based 
on early
signals extracted from tweets. Example applications include detecting activity
planning~\cite{Becker2012} and forecasting future events such as civil
unrest~\cite{2623373} and upcoming threats to national
airports~\cite{312122121}. In our work, we follow a similar intuition to detect signals for major security attacks. The key novelty in our approach,
different from these works, is the
need for a typed query expansion strategy that provides both focused
results and aids in extracting key indicators underlying the
cyber-attack.

\begin{table*}[htpb]
\small
    \centering 
    \begin{tabular}{cP{1.2cm}P{1.9cm}P{1.70cm}P{1.70cm}cccP{1.85cm}}
        \toprule
        \multirow{2}{*}{} & \multicolumn{3}{c}{Method} & 
        \multicolumn{3}{c}{Event} & \multirow{2}{*}[0.65em]{Goal} &
        \multirow{2}{*}[0.65em]{Data}\\
        \cmidrule{2-4}\cmidrule{5-7}
                                 & Unsupervised & Keyword Expansion & 
        Information Extraction  & Characterize Event
        & Type & Detection & &\\
        \midrule
         \cite{191002} & 
        \ding{53} & \ding{53} &  \ding{53} %
        & \ding{53} & \ding{53} & \checkmark
        & Cyberattacks & Network Data \\
       \midrule
        \cite{Ritter:2015:WSE} &
        \ding{53}  & \checkmark & \checkmark
        & \ding{53} &  \checkmark &  \checkmark
        & Cyberattacks & Twitter \\
         \midrule
        \cite{Michaelcss16} &
        \ding{53} & \ding{53} & \ding{53}
        & \ding{53} & \ding{53} & \ding{53}
        & Cyberattacks & WINE \\
        \midrule
        \cite{zhu_dumitras_2016} & 
        \ding{53} & \ding{53} & \checkmark%
        & \ding{53} & \ding{53} & \ding{53}
        & Malware & Papers \\
        \midrule
        \cite{Kwon:2015:DEI:2810103.2813724} &
        \ding{53} & \ding{53} & \ding{53}%
        & \ding{53} & \ding{53} & \ding{53}
        & Malware & WINE \\        
        \midrule
        \cite{191006} &
        \ding{53} & \ding{53} & \checkmark
        & \ding{53} & \ding{53} & \checkmark
        & Vulnerability & Twitter \\
        \midrule
        \cite{Ding:2012:ISC:2339530.2339670} &
        \checkmark& \ding{53} & \ding{53}%
        & \ding{53} & \ding{53} & \ding{53}
        & Intrusion & Network Data \\
        \midrule
        \cite{Noble:2003:GAD:956750.956831} &
        \checkmark& \ding{53} & \ding{53}%
        & \ding{53} & \ding{53} & \ding{53}
        & Intrusion &  Network Data\\
        \midrule
        \cite{4476697} &
        \checkmark& \ding{53} & \ding{53}%
        & \ding{53} & \ding{53} & \ding{53}
        & Intrusion &  Network Data \\
        \midrule
        \cite{Davis:2011:DAG:2063576.2063749} &
        \checkmark& \ding{53} & \ding{53}%
        & \ding{53} & \ding{53} & \ding{53}
        & Insider & Access Log \\
        \midrule
        \cite{Liao:2016:AIG:2976749.2978315} &
        \checkmark& \ding{53} & \checkmark%
        & \ding{53} & \ding{53} & \ding{53}
        & IOC & Tech Blogs \\
        \midrule
        Ours &
        \checkmark& \checkmark& \checkmark%
        & \checkmark& \checkmark& \checkmark%
        & Cyberattacks & Twitter \\
        \bottomrule
    \end{tabular}
    \caption{Related work.}\label{tab:related_work}
\end{table*}

\section{Conclusion}
We have demonstrated an unsupervised approach to extract and
encode cyber-attacks reported and discussed in social media. 
We have motivated the need for a careful template-driven query
expansion strategy, and how the use of dependency parse trees
and word embeddings supports event extraction.
Given the widespread prevalence of cyber-attacks, tools such as
presented here are crucial to providing situational awareness on
an ongoing basis. 
Future work is aimed at broadening the class of attacks that the system
is geared to as well as at modeling sequential dependencies in cyber-attacks.
This will aid in capturing characteristics such as increased
prevalence of attacks on specific institutions or countries during particular
time periods.

\bibliographystyle{abbrv}

\end{document}